\documentclass[conference]{IEEEtran}
\IEEEoverridecommandlockouts
\pdfoutput=1

\usepackage{cite}
\usepackage{amsmath,amssymb,amsfonts}
\usepackage{algorithmic}
\usepackage{graphicx}
\usepackage{textcomp}
\usepackage{xcolor}
\def\BibTeX{{\rm B\kern-.05em{\sc i\kern-.025em b}\kern-.08em
    T\kern-.1667em\lower.7ex\hbox{E}\kern-.125emX}}

\usepackage[ruled,linesnumbered]{algorithm2e}
\usepackage{multirow}
\usepackage{subfig}
\usepackage{graphicx}
\usepackage{color}

\newlength\mylen
\newcommand\myinput[1]{%
  \settowidth\mylen{\KwIn{}}%
  \setlength\hangindent{\mylen}%
  \hspace*{\mylen}#1\\}
  
\let\oldnl\nl
\newcommand{\nonl}{\renewcommand{\nl}{\let\nl\oldnl}}

\newcommand{\deceit}{\textsl{DECEIT}\xspace}

\begin{document}
%

\title{Metamorphic Detection of Repackaged Malware \\
{\footnotesize 
\thanks{The Programming Systems Laboratory is supported in part by NSF CNS-1563555 and CCF-1815494.}
}}



\author{\IEEEauthorblockN{Shirish Singh}
\IEEEauthorblockA{\textit{Department of Computer Science} \\
\textit{Columbia University}\\
New York NY, USA \\
shirish@cs.columbia.edu}
\and
\IEEEauthorblockN{Gail Kaiser}
\IEEEauthorblockA{\textit{Department of Computer Science} \\
\textit{Columbia University}\\
New York NY, USA \\
kaiser@cs.columbia.edu}
}



\maketitle

\begin{abstract}


Machine learning-based malware detection systems are often vulnerable to evasion attacks, in which a malware developer manipulates their malicious software such that it is misclassified as benign. 
Such software hides some properties of the real class or adopts some properties of a different class by applying small perturbations.
A special case of evasive malware hides by repackaging a bonafide benign mobile app to contain malware in addition to the original functionality of the app, thus retaining most of the benign properties of the original app.   
We present a novel malware detection system based on metamorphic testing principles that can detect such benign-seeming malware apps. 
We apply metamorphic testing to the feature representation of the mobile app, rather than to the app itself.  
That is, the source input is the original feature vector for the app and the derived input is that vector with selected features removed.  
If the app was originally classified benign, and is indeed benign, the output for the source and derived inputs should be the same class, i.e., benign, but if they differ, then the app is exposed as (likely) malware. Malware apps originally classified as malware should retain that classification, since only features prevalent in benign apps are removed.
This approach enables the machine learning model to classify repackaged malware with reasonably few false negatives and false positives. 
Our training pipeline is simpler than many existing ML-based malware detection methods, as the network is trained end-to-end to jointly learn appropriate features and to perform classification. 
We pre-trained our classifier model on 3 million apps collected from the widely-used AndroZoo dataset.\footnote{https://androzoo.uni.lu/} 
We perform an extensive study on other publicly available datasets to show our approach's effectiveness in detecting repackaged malware with more than 94\% accuracy, 0.98 precision, 0.95 recall, and 0.96 F1 score.

\end{abstract}

\begin{IEEEkeywords}
machine learning, malware detection, repackaged malware, mobile apps
\end{IEEEkeywords}


%

\section{Introduction}


Mobile devices are prevalent in our daily lives.
Android mobile devices continue to dominate the global mobile market, with 86.1\% market share in 2019, according to the information published by IDC~\cite{IDCSmart5:online}. 
Android malware is a persistent threat to billions of users around the world: A 2015 survey~\cite{OutofPoc54:online} reports that millions of malicious applications have been found on mobile phones, and 96\% of them aim at the Android system. Thus, malware detection for mobile apps has been an active research topic in recent years. Researchers often employ static or dynamic analysis techniques to discover malware, many of which apply machine learning to the data extracted from the analysis techniques to make classify apps as malware or benign. 
However, limited work has been done on machine learning approaches to predicting malware that employs 
\emph{repackaging}~\cite{li2019rebooting}. 

In this paper, we present a novel technique, \deceit (\textbf{DE}te\textbf{C}ting repackaged malware with m\textbf{E}tamorph\textbf{I}c \textbf{T}esting), to address the challenge of detecting repackaged malware. Our approach, based on metamorphic testing (MT) principles,
automatically detects malware that leverages repackaging without needing to know the true output label for every app (benign or malware). 
Our technique does not require re-training a machine learning model, only metamorphic relations that selectively reduce feature vectors.

\begin{figure*}[!htp]
\centering
\begin{tabular}{ccc}
\includegraphics[width=.31\textwidth]{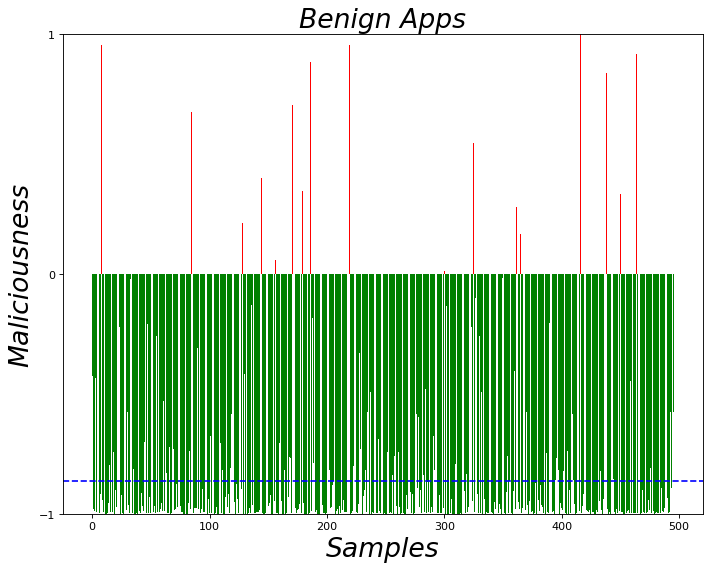} &
\includegraphics[width=.31\textwidth]{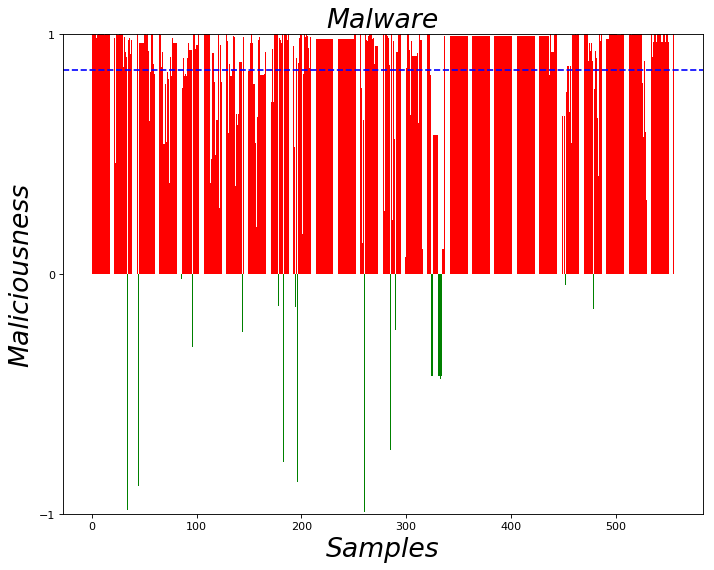} & 
\includegraphics[width=.31\textwidth]{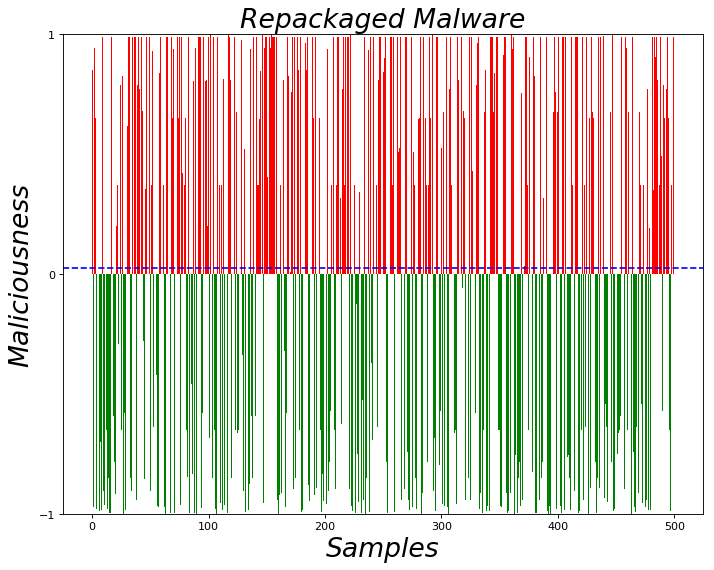} \\
\textbf{(a) Benign Apps}  & \textbf{(b) Drebin Malware} & \textbf{(c) Repackaged Malware}\\
\end{tabular}
\medskip
\vspace*{-10pt}
\caption{Red and green bars represent apps identified as malware and benign, respectively, and their length refers to the prediction probability. The blue dotted line shows the average prediction probability of that set. We retrieved the maliciousness of each app based on the classifier's probability. Figure (a) shows the maliciousness of most benign apps is below the set threshold. Figure (b) shows that the maliciousness of many malware apps from the Drebin dataset are above the threshold. Figure (c) represents the maliciousness of repackaged malware samples. 
}
\label{fig:malware_evolution}
\vspace{-10pt}
\end{figure*}

We stress that this work stretches MT principles in a novel direction: The conventional MT concept of a source input/output with a followup derived input and its predicted vs. actual outputs is not employed directly. That is, given an Android APK as input, a machine learning classifier, and a binary class C as output (either benign or malware), we do not use a metamorphic relation (MR) to derive a new APK' as followup input to the classifier. (Android apps are distributed as APKs.  "APK" stands for Android Package and is the Android equivalent of a Java JAR file.)  Instead, we consider the same input Android APK and employ an MR to derive a followup feature vector V' from that APK's original feature vector V.  
The MRs do not change the values of any of the original features, they simply remove the selected features from consideration by the classifier. Different MRs would thus remove different subsets of the features -- but carefully selected, not at random, just like conventional MRs do not (usually) arbitrarily change the source inputs, instead the derived inputs are carefully crafted to enable a meaningful output prediction. In our \deceit approach, the output class is always predicted to be the same as the original output C.  Divergence between the prediction of the followup output C' and the actual followup output C'' is interpreted as a limitation of the classifier in its ability to distinguish benign from malware APKs.  In particular, its original classification of the APK is suspect and probably wrong.  



\subsection{Motivation}


The impact of a malware campaign on Android can be enormous, which was clearly observed in the case of the Judy malware attack that was pushed through the Google Play stores and infected 36 million users worldwide~\cite{TheJudyM72:online}. Attacker strategies are changing and current defenses find it hard to cope with new malware variants, which makes Android malware detection a challenging problem. 
Machine learning approaches are widely used to detect malware. However, machine learning algorithms can be easily spoofed by carefully perturbing the inputs presented to the detector.
Many studies have been done on such ``adversarial'' Android malware~\cite{chen2019android, grosse2017statistical, Liu_2019}. 
Repackaged malware is perhaps one of the simplest approaches in this regard, since the malware inherits most of the features of the benign app in which it is embedded.
Grosse et al. \cite{2016arXiv160604435G} demonstrated that a highly accurate deep learning Android malware classifier can be evaded by simply adding less than 20 features to the Android app's manifest file.

To demonstrate the evolution of mobile malware, we trained a simple single-layer neural network classifier on malware and benign apps from the Drebin dataset~\cite{arp2014drebin} containing malware and benign app samples from before 2014. We tested the model on three categories of apps: benign apps (randomly sampled), malicious apps (subset) from the Drebin dataset, and repackaged apps \cite{li2019rebooting}. We observed that repackaged malware exhibit
many characteristic features of benign apps, such as common API and library usage, which consequently led the model to misclassify these malware as benign. Figure~\ref{fig:malware_evolution} demonstrates the classification probabilities of the model. The maliciousness score is the model's prediction probability, and it lies in the range of -1 to 1, with -1 representing benign and 1 representing malware. Figure~\ref{fig:malware_evolution} (a) shows the classification probability of 500 randomly sampled benign apps. 
Figure~\ref{fig:malware_evolution} (b) and (c) show the stark contrast between the classification of Drebin malware and repackaged malware samples, respectively. Figure~\ref{fig:malware_evolution} (c) shows mixed prediction on repackaged malware. Such differences in the prediction probabilities are attributed to evasion attacks wherein the malware exhibits a considerable amount of benign features such that the malicious characteristics are hidden from a naive classifier.



The difference in the prediction probabilities is evident in the case of repackaged malware, which poses a serious threat to the Android ecosystem as it deprives app developers of their benefits, contributes to spreading malware on users' devices, and increases the workload of market maintainers~\cite{li2019rebooting}. In the repackage malware dataset~\cite{li2019rebooting}, we observed that repackaged malware apps and benign apps share more than 80\% of their features, which makes detection a challenging problem. In this work, our primary objective is to detect repackaged malware.
Our novel framework tackles the problem of repackaged malware efficiently. Many state-of-the-art approaches, e.g., \cite{hindroid, KARBAB2018S48}, have reported high malware detection rates on closed datasets, which are difficult to verify. We used AndroZoo (open dataset) \cite{androzoo}, Drebin dataset (Publicly available) \cite{arp2014drebin} and Repackaged apps (open dataset) \cite{li2019rebooting} to demonstrate the effectiveness of our framework.






\subsection{Contributions}

In this paper we address the aforementioned challenges while building a robust framework for detecting repackaged malware that circumvent state-of-the-art machine learning-based detection techniques. The contributions of this paper are:

\begin{enumerate}
  \item We present an adaptation of metamorphic testing principles to machine learning models, to apply metamorphic relations to feature vectors rather than the underlying information sources represented by the features.
  
  \item We apply this idea to mobile malware detection, to identify repackaged malware. 

  \item We conduct experiments on real-world repackaged malware samples, and the results show that our novel technique successfully detects malware with more than 94.56\% accuracy, 0.98 precision, 0.95 recall, and 0.96 F1 score.
\end{enumerate}

Our simple approach is beneficial for several reasons. First, it increases the difficulty for attackers to exploit the vulnerabilities of a model. Second, our adversary-resistant technique maintains desirable classification performance while requiring only minimal modification to the classification process. Ultimately, while this work is primarily motivated by the need to detect repackaged malware, it should be noted that our adaptation of metamorphic principles is general and could potentially be adapted to other ML applications.
Our open-source implementation, models, and complete training \& testing datasets are available at https://github.com/Programming-Systems-Lab/DECEIT. 


Section~\ref{sec:background} explains background on malware, neural networks and model interpretability. Section~\ref{sec:related_work} discusses related work followed by Section \ref{sec:meta_relation} discussing the metamorphic relations used in our study. Section ~\ref{sec:system_overview} provides a detailed description of our framework.
Section~\ref{sec:dataset} describes our experimental setting, data acquisition methodology, feature extraction, and dataset compilation.
Section~\ref{sec:results} presents our evaluation scheme and experimental results. Threats to validity and limitations are presented in Section \ref{sec:threats} followed by our conclusions.

\begin{figure*}[!ht]
    \centering
    \includegraphics[width=\textwidth]{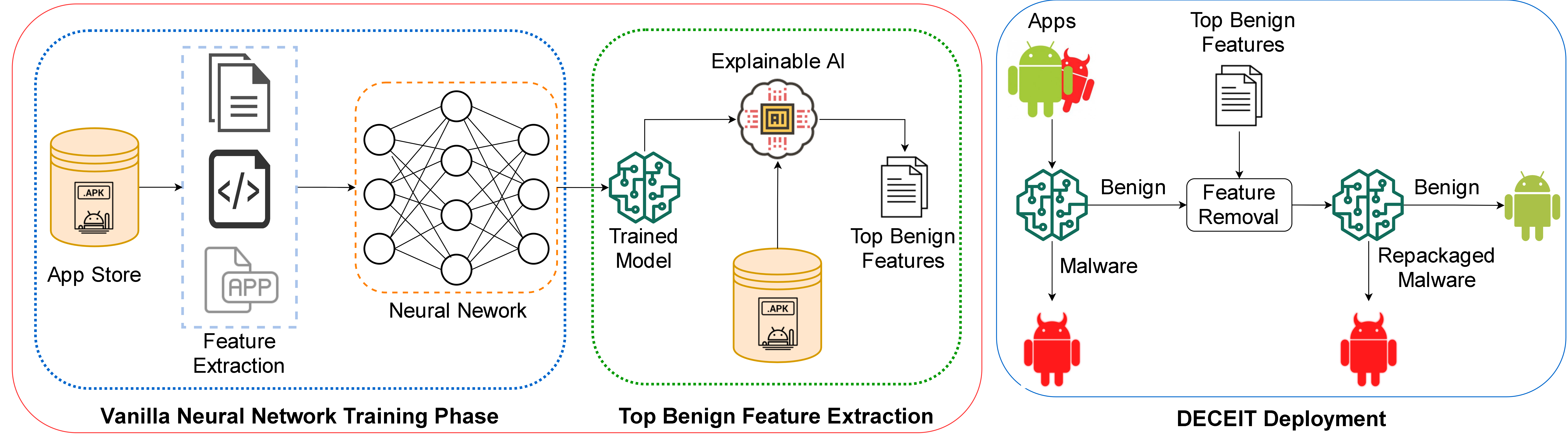}
    \caption{\deceit Overview}
    \label{fig:overview}
    \vspace{-10pt}
\end{figure*}

\section{Background}
\label{sec:background}

\subsection{Repackaged Apps}

Repackaging refers to the core process of unpacking a software package, then repackaging it after a probable modification to the decompiled code and/or to other resource files (e.g., libraries, Permissions). Two apps are considered repackaged app pair if they share at least 80\% of the code (i.e., code similarity exceeds 80\%), and are signed by different developer(s) \cite{li2019rebooting}. The repackaged apps include malicious payload and execute malicious code. Sharing 80\% of the application structure and code with benign apps makes detecting repackaged malware incredibly difficult.

\subsection{Neural Networks}

Neural networks are a kind of machine learning models that rely on computational units -- neurons -- organized in layers to learn patterns in the underlying data. A ``feed-forward'' neural network consists of a sequence of neuron layers. Each layer of the network produces an output, which is used as an input to the next layer. The neurons in sequential layers are connected to each other through weighted edges. Neurons apply an activation function to the input values and pass the results to the neurons in the subsequent layer.

In a classification task, a feed-forward neural network takes a feature vector as an input to train the model. The feature vector can be represented as a vector of numbers representing attributes of the input sample. For example, an Android app can be represented using a feature vector containing binary values that capture the presence/absence of components in Android apps. A neural network classifier learns the decision boundary between the classes. A good classifier categorizes the identical samples correctly, regardless of the underlying feature representation. Networks with multiple intermediate hidden layers are called Deep Neural Networks (DNNs).



\subsection{Model Interpretability}


Model interpretability gives machine learning models the ability to explain or to present their behaviors in understandable terms to humans \cite{doshivelez2017rigorous}. Such explanations can help better understand the data and why a model might fail, and eventually increase the system safety \cite{10.1145/3359786}.

Local Interpretable Model-agnostic Explanations (LIME) is a popular method for explaining predictions of machine learning classifiers proposed by Ribeiro et al. \cite{ribeiro2016should}. The 
intuition behind 
LIME is to learn the response of the underlying interpretable model to the perturbation of the input and see how the prediction changes \cite{LocalInt26:online}. The model's decision function is a nonlinear function. LIME samples instances and gets predictions using the decision function, and weighs them by the proximity to the instance being explained. At last, a linear model is learned to locally approximate the decision boundary in the vicinity of the explained instance. The purpose of using LIME in our work is to identify features that are important for classification so that we can have a better understanding of feature distributions across malware and benign apps.

\section{Related Work}
\label{sec:related_work}


Malware detection systems have been extensively studied \cite{arp2014drebin, ALZAYLAEE2020101663, 10.1145/3372297.3417291, 7412132, singh2015artificial}, especially after proliferation of smartphones. Traditional malware detection techniques relied on maintaining a list of malicious apps derived by computing signature of apps \cite{idika2007survey}. Such techniques perform poorly on detecting new (unknown) malware. However, advances in machine learning have enabled researchers to build intelligent malware detection systems capable of detecting newer malware. 
Machine learning based techniques primarily rely on features generated from static analysis and dynamic analysis of apps. Though effective, ML based techniques have been shown to be susceptible to adversarial attacks \cite{10.1145/3052973.3053009}.

Prior work have explored defenses against adversarial attacks:
Adversarial training \cite{42503} increases robustness by augmenting training data with adversarial examples, however, generating valid samples is computationally intensive and may break the inherent application constraints.
Other works, such as, Incer et al. \cite{10.1145/3180445.3180449} demonstrated the effectiveness of using monotonic classifiers, where adding features can only increase the decision score. Such classifiers require significant time towards manual feature selection. Adversarial robustness can be further improved by using an ensemble of classifiers for prediction \cite{Li_2020}, adding noise to the training data \cite{2015arXiv151106306J},
dimensionality reduction \cite{Xu_2018}, and removing adversarial examples after detection \cite{2017arXiv170204267H},
In contrast, \deceit does not require re-training the model or access to the training dataset to apply its metamorphic relations to identify (likely) repackaged malware.


Murphy et al.~\cite{MLMetamorphicTesting} was probably the first work to apply metamorphic testing to machine learning software; they identified several general types of metamorphic relations (MRs) that apply to most machine learning applications. 
Xie et al.~\cite{5381489,XIE2011544} studied the application of metamorphic testing specifically to ML classifiers. Tian et al.~\cite{10.1145/3180155.3180220} applied metamorphic testing to DNNs.  Mekala et al. \cite{10.1109/MET.2019.00016} proposed using metamorphic testing principles to automatically detect adversarial attacks on image classifiers, by applying metamorphic relations based on distance ratio preserving affine image transformations.



\section{\deceit's Metamorphic Relations}
\label{sec:meta_relation}

Machine learning models rely on feature representation of the data to make a classification. Here we used binary features to represent the presence/absence of components in Android apps. Hence, the model learns how the existence or absence of a feature impacts each class. Based on this property, we investigated a generic template for metamorphic relations concerned with classifying examples as repackaged or clean benign apps: ``Removal of Top-Benign features''. Given an app's feature representation, $f_{app}$, and predicted classification from the model, $C$, removing several of the top benign features from $f_{app}$ should not affect the prediction of the model. \deceit relies on this property to detect malicious apps.

\begin{figure}[ht]
  \centering
    \includegraphics[width=\linewidth]{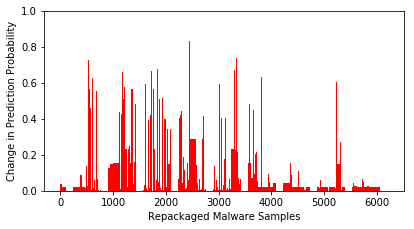}
    \caption{Change in prediction probability of vanilla neural network after removing top 6 benign features from repackaged malware. We observe that several apps see a change in the detection probability. Higher the probability change, more likely the app is a malware.}
    \label{fig:probability_change_malware}
\end{figure}

\begin{figure}[ht]
  \centering
    \includegraphics[width=\linewidth]{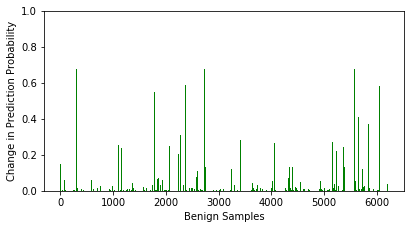}
    \caption{Change in prediction probability of vanilla neural network after removing top 6 benign features from benign apps. We observe that fewer apps see change in the prediction probability.}
    \label{fig:probability_change_benign}
\end{figure}

The classification probability of an app should change if the app's feature representation is manipulated. We hypothesize that removing selected (top) benign features should result in a greater change in probability towards malware classification in the case of repackaged malware. However, for truly benign apps, removing such features should result in more limited changes to the classification probability. This relatively small change in the classification probability of the benign apps can be attributed to the existence of other benign features that influence the prediction. However, in malware, after removing the benign features, the malicious features become more prominent, allowing the model to detect the malware. Figures \ref{fig:probability_change_malware} and \ref{fig:probability_change_benign} show the probability change when removing top-6 benign features from benign and malware apps. Greater the probability change, the more likely the app is malware. We observe that when removing the top benign features from the repackaged malware dataset, the change in the detection probability can be seen in many apps as compared to the benign apps dataset. The following two equations are used to transform the feature vector and make classification, respectively.

\begin{equation}
    f'_{app} = f_{app} - f_{top\_benign\_features}
\end{equation}
\vspace{-10pt}
\begin{equation}
  Classification =
    \begin{cases}
      Malware, & P(f'_{app}) > \delta \\
      Benign, & \text{otherwise}
    \end{cases}       
\end{equation}

where $f_{app}$ is the feature vector representing an app. $f_{top\_benign\_features}$ represents top benign features, then, $f'_{app}$ is the feature vector of the app after removal of top being features. $P(x)$ represents the classifier's prediction probability, and $\delta$ is the decision threshold. The higher the prediction probability, the more likely the app is malware. Oftentimes, in a binary classification problem, the default value for the threshold is set to 0.5, where all values equal or greater than the threshold are mapped to one class, and all other values are mapped to another class. Since we are also performing binary classification, we use 0.5 as the decision threshold ($\delta$). 

\section{\deceit Overview}
\label{sec:system_overview}

\begin{figure}[ht]
  \centering
    \includegraphics[width=0.9\linewidth]{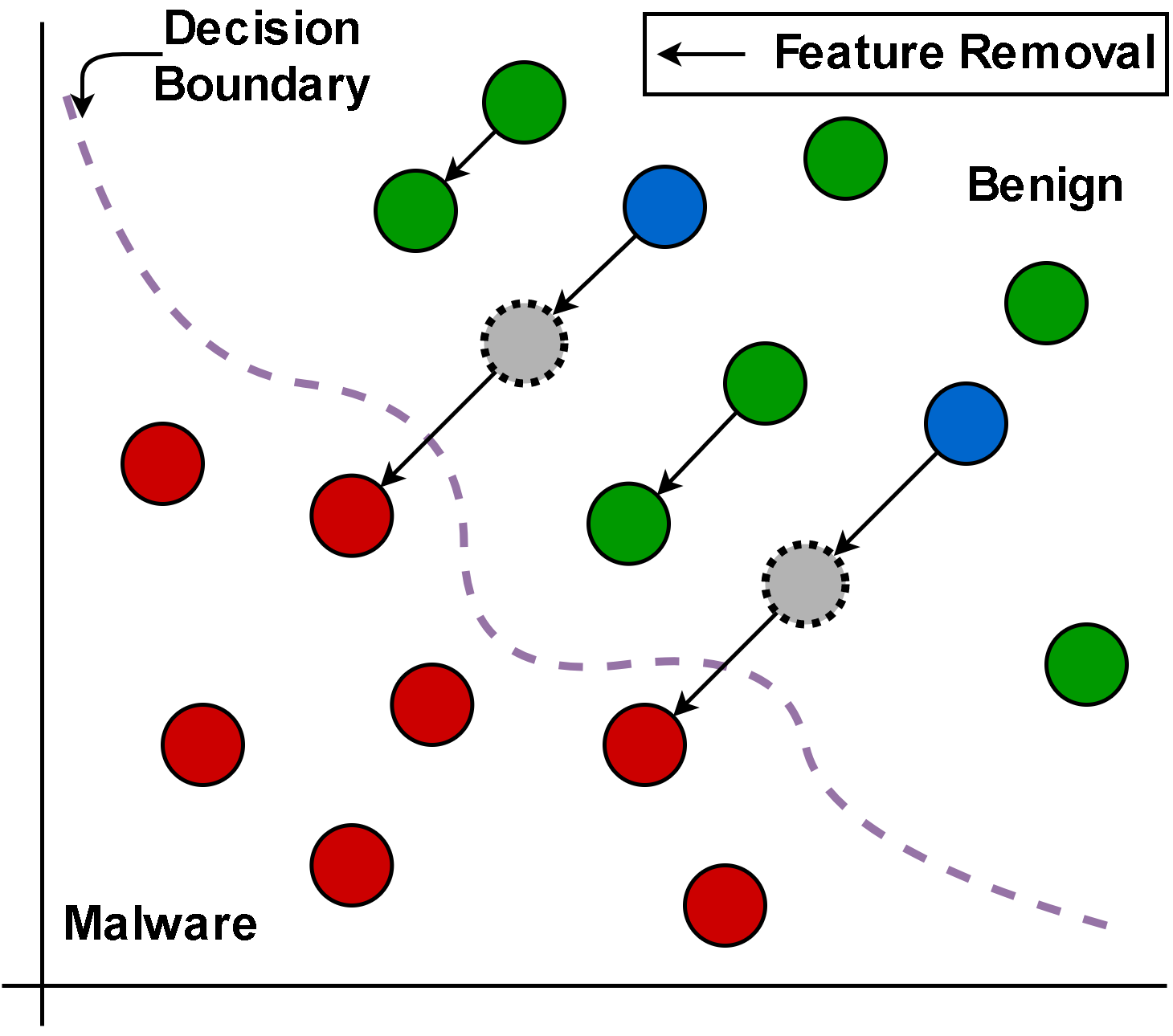}
    \caption{The red and green dots refer to malicious and benign apps, respectively. The blue dots represent evasive repackaged malware misclassified as benign. Upon selective nullification of top benign features, the blue dots shift towards the decision boundary and finally turn red, exposing their maliciousness. The arrows direct toward the transformation process.}
    \label{fig:deceit_example}
\end{figure}

\deceit relies on two prime components: a model and its interpretability to identify the relevant features and perform the classification. The fundamental concept is to observe the model's prediction probability change after exposing the malicious features of apps by removing top-benign features. The remaining features are then used to determine the probability of the application being malware. Figure \ref{fig:overview} shows a high-level overview of \deceit. We introduce selective feature nullification that allows us to unravel the adversarial features, hidden among the benign features of the application, used to circumvent detection on conventional detection systems. Figure \ref{fig:deceit_example} shows how removing benign features from evasive malware leads to correct malware classification. Our approach is fundamentally different from simply changing the threshold of a classifier. Since each feature has a different impact on the classification probability, removing the same feature from two different apps results in different changes in the apps' prediction probabilities.

We train a simple vanilla neural network model on the training data derived from the AndroZoo dataset (see section \ref{sec:dataset}). Once the model is trained, we use LIME to extract, categorize, and rank the features used by the model to classify the apps. These features define the prediction outcome of the classifier. The probability distribution of features is learned for benign and malware, and we split it into two: benign set and malware set. When Benign features are removed, the detection score of malicious apps increases. This phenomenon continues until a threshold $t$ is reached, with $t$ determined empirically. Algorithm \ref{algo:deceit} shows how LIME is used to aggregate the top benign features and make predictions.

\begin{algorithm}
 \KwData{ Training Feature Matrix $X_{train}$}
 \nonl\myinput{Dev sample $X_{dev}$}
 \nonl\myinput{Test sample $X_{test}$}
 \KwResult{Classification of $X_{test}$}
 Train a classifier \textit{clf}\ with prediction function $P(X_{sample})$\;
 Initialize explainable AI (LIME) function $E(X_{sample},\ X_{train},\ clf)$\;
 Initialize\ $f_{top\_benign}$ as list of top-benign features\;
 \For{$X_{sample}\ \epsilon\ X_{dev}$}{
   $f_{X_{sample}}$ = $E(X_{sample},\ clf)$ \;
   $f_{top\_benign}\ +=\ f_{X_{sample}}$ \;
 }
 \eIf{P ($X_{test}$\ =\ malware)\ $\ge$\ $\delta$}{
   Prediction is Malware\;
   }{
   Remove $f_{benign}$ from $X_{test}$ to get $X'_{test}$\;
   \uIf{P ($X'_{test}$\ =\ malware)\ $\ge$\ $\delta$}{
    Prediction is Malware\;
   } \Else{
    Prediction is Benign\;
   }
  }
 \caption{Algorithm}
 \label{algo:deceit}
\end{algorithm}

\textbf{Top-benign features}: We used LIME \cite{ribeiro2016should} to first explain the predictions of the classifier and then extract the features indicative of benign characteristics for every prediction in the development dataset (see section \ref{sec:dataset} for dataset distribution). We counted the number of occurrences of each feature in each prediction and ranked the features based on the count obtained. Table \ref{tab:top_features} shows the top 20 benign features, ranked by counts.

\begin{table}[!ht]
  \centering
  \begin{tabular}{l}
  \hline
  \textbf{Top-20 Benign Features} \\ \hline
  Class: java.lang.ClassLoader \\ 
 Class: android.content.ContentResolver \\ 
 Class: android.content.ContentProvider \\ 
 Class: android.net.ConnectivityManager \\ 
 Class: android.telephony.TelephonyManager \\ 
 Intent: action.BATTERY\_CHANGED \\ 
 Intent: action.PACKAGE\_REPLACED \\ 
 Intent: extra.CONTENT\_ANNOTATIONS \\ 
 Intent: action.ACTION\_SHUTDOWN \\ 
 Class: android.content.pm.PackageInstaller \\ 
 Class: android.net.http.SslCertificate \\ 
 Permission: MANAGE\_OWN\_CALLS \\ 
 Package: org.apache.http.params \\ 
 Permission: CAMERA \\ 
 Intent: action.SEARCH \\ 
 Intent: action.ACTION\_POWER\_CONNECTED \\ 
 Permission: CALL\_PHONE \\ 
 Intent: action.EDIT \\ 
 Intent: category.MONKEY \\ 
 Intent: action.MANAGED\_PROFILE\_UNLOCKED \\
  \end{tabular}
  \caption{Top-20 ranked benign features extracted from the development set}
  \label{tab:top_features}
\end{table}


\section{Dataset}
\label{sec:dataset}

\textbf{App Vetting} VirusTotal~\cite{VirusTot56:online} is a popular online platform that aggregates many Anti-Virus (AV) products and online scan engines to check for malware. Salem et al. \cite{DBLP:journals/corr/abs-1903-10560} showed that the scan results obtained from VirusTotal are continuously changing with time and, hence, should not be taken for granted unless they are up-to-date. Secondly, we found that such VirusTotal results –- regardless of their freshness -– can significantly alter the composition of a dataset (i.e., which apps are malicious and which are benign), depending on the scheme adopted to label apps in a dataset. Such schemes depend on how the VirusTotal composite score is converted to a single binary label of malware or benign \cite{DBLP:journals/corr/abs-1903-10560}. We labeled an app as benign if all the AV scanners detect no suspicious behavior, and as malware if the app was tagged by more than 10 scanners.
All other samples were disregarded.

\textbf{Training Dataset} We collected the 3,226,927 APKs from the AndroZoo dataset  \cite{androzoo}. 
Each APK was vetted by VirusTotal (VT).
Post vetting, we used 2,572,426 benign apps and 418,973 malware, for a total training dataset of 2,991,399 apps. We used 90\% of the data to train the model and 10\% to tune the hyper-parameters.

\textbf{Testing Dataset} To evaluate the performance of our framework we use repackaged Android malware dataset RePack \cite{li2019rebooting} consisting of 15,297 repackaged Android apps, from which we selected the 6,209 apps tagged by at least ten VT scanners as malware. We also used benign Google Play apps. We sampled 6,209 apps, tagged as benign by VT, to represent the different categories of Android apps. In total we tested the performance on 12,418 apps.

\subsection{Feature Extraction}

All features were extracted using static analysis on the apps. Similar to Drebin \cite{arp2014drebin}, we focused on the manifest and the disassembled dex code of the apps to extract features. A linear sweep over APK can obtain these features. We extracted static features from applications' APK file using Androwarn \cite{maaaazan34:online}, which uses Androguard \cite{androguard} to reverse engineer the APK file. We embed apps in a binary feature space which captures the presence/absence of components in Android apps. Each application in our data-set was represented by a feature vector of 693 features, which are broadly sub-divided into six categories: 
Permissions, Packages, Hardware, Intents, Classes, and Leaks. In addition to these features extracted from the APK file, we check for code obfuscation within application's code and various malicious behaviour analysis produced by Androwarn, adding another feature to make 694.

\section{Results}
\label{sec:results}

\textbf{Vanilla Network Architecture}: Our network consists of input layer (containing 694 neurons corresponding to the 694 features), hidden layer(ReLu activation), and a softmax output layer. 
We train the network on the Androzoo dataset and use it as a baseline for the experiments. Our classifier achieves up to 98\% accuracy with minimal effort for hyper-parameter selection. This performance matches existing malware detection systems that rely on static features.

\subsection{Performance on Repackaged Apps}
\label{sec:preformance_repackaged_apps}

We collected repackaged apps from Re-Pack repository \cite{li2019rebooting}. This dataset consists of 15,297 original-repackaged apps. 
Not all repackaged apps are malicious; some can be cloned or plagiarised apps. We were interested in a sub-set of repackaged apps: piggybacked or camouflaged \cite{10.1109/TIFS.2017.2656460}. These apps are repackaged to include malicious payload and execute malicious code. Sharing 80\% of the application structure and code with benign apps makes detecting repackaged malware incredibly difficult.

\begin{figure}[!ht]
  \centering
    \includegraphics[width=\linewidth]{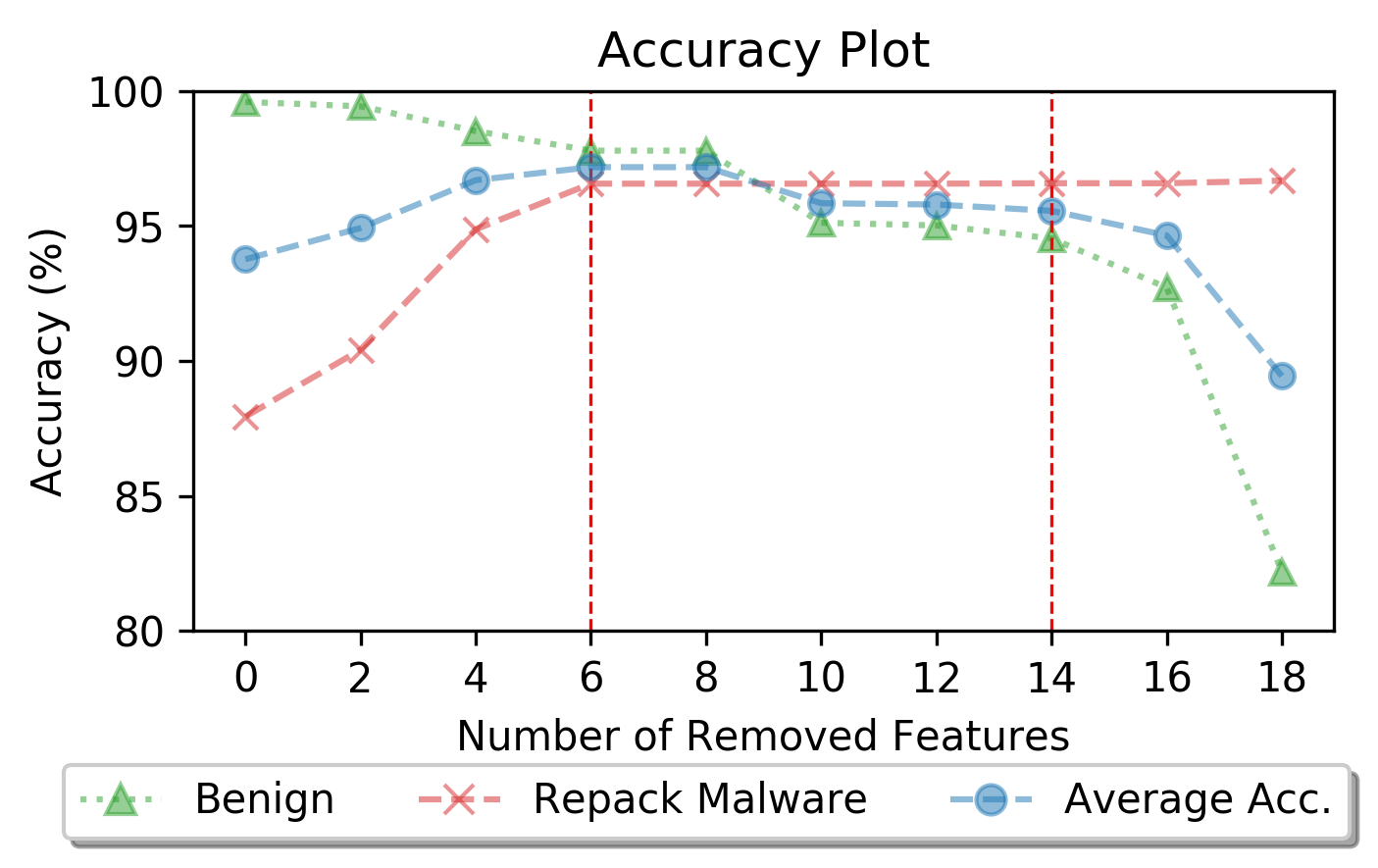}
    \caption{The accuracy vs. the number of features removed plot demonstrates how removing ranked benign features augment the detection accuracy of repackaged malware apps while consistently detecting benign apps. However, the average accuracy degrades slightly after removing six features and then more significantly because of increment in false-positive rates.}
    \label{fig:repackaged_result}
    \vspace{-10pt}
\end{figure}

We collected VirusTotal reports of all apps to get the ground truth. We classified an app as malware if more than ten detector identified the app as malware. Ultimately, we had 6,209 malware.
The performance on malicious repackaged app detection is shown in figures \ref{fig:repackaged_result}.
The detection accuracy of the vanilla network is 87.82\% with no features removed. As the features are removed, the detection rate goes up consistently to a point. After removing 2 features, the increment in the performance is negligible. This behavior can be attributed to the fact that repackaged apps share about 80\% of the code-base with the benign apps, as stated above, and only 20\% of the app is unique. After removing 6 features, we observe a detection rate of 94.56\% (An increment of 6.74\%).

\subsection{Performance on Benign Apps}

Vanilla model has a detection rate of 99.59\%. However, after removing top 6 benign features, we achieve the highest malware detection rate without compromising significantly on benign apps' detection rate; in particular, the accuracy for benign apps reduces slightly to 97.77\%. As we can see, adopting \deceit causes a 1.82\% drop in detection rate Of benign apps. The benefit of our approach is its security against repackaging attacks. Thus, we interpret the 1.8\% difference as the cost
of security for malware classification. We also observe that after removing top-14 features, the average accuracy degrades significantly because of high false-positive rates. 

Table \ref{tab:performance} summarizes the results of vanilla NN and DECEIT.

\begin{table}[!ht]
    \centering
    \begin{tabular}{c|c|c|c|c}
        \textbf{Model} & \textbf{Precision} & \textbf{Recall} & \textbf{F1} & \textbf{Accuracy} \\ \hline
        Vanilla NN & 0.99 & 0.88 & 0.93 & 93.71\% \\
        DECEIT & 0.98 & 0.95 & 0.96 & 96.17\% \\
    \end{tabular}
    \caption{Performance Comparison}
    \label{tab:performance}
\end{table}

\section{Threats to Validity and Limitations}
\label{sec:threats}


We avoided bias in training and testing by leveraging APK datasets collected by others, treating as malware only those APKs tagged by at least ten VT scanners and treating as benign only those APKs not tagged by any VT scanners as malware.

We used LIME \cite{ribeiro2016should} to retrieve the top benign features from the development dataset. LIME is designed for approximating locally in the neighborhood of the prediction we want to explain. Thus the feature distribution can change if new apps are sampled from the dataset to extract benign features. This change could potentially impact the choice of which features to eliminate and the results obtained after applying feature elimination. An alternate model-explainability tool such as SHAP \cite{NIPS2017_7062} could potentially give better results at the cost of computation time. SHAP can guarantee properties like consistency and local accuracy. Another potential threat is the choice of neural architecture. In this work, we utilized a simple vanilla neural network with limited sets of features. More sophisticated network architectures could yield potentially better results on the same dataset. 

We observed that while the detection accuracy of malware increases consistently as we remove the top benign features, the accuracy on benign apps reduces marginally at first and then more significantly. In our studies, we also observed that the optimal number of features that can be removed without compromising the accuracy of the model is in the range of 2 to 10, as seen in section \ref{sec:results}, figure \ref{fig:repackaged_result}. We do not address the problem of model aging \cite{10.1145/3372297.3417291}. However, our approach can be augmented by applying such techniques.

Malware detection is a time series problem. Newer technologies expose new attack surfaces that can be exploited by novel threats. 
We rely on features extracted from the application packages. The performance is dependent on the types of features used to represent the data. Given the nature of malware detection, the model requires re-training as newer APIs are added to the Android operating system and to adapt to newer malware trends. 

\section{Conclusion}

We present a novel technique based on metamorphic testing principles, \deceit, to expose repackaged malware apps that utilize benign features to hide their true nature from the machine learning classifier. Our approach is quite different from previous work that applies metamorphic testing to find bugs in machine learning software.  We are not looking for bugs, per se.  We apply metamorphic testing to the feature representation of the mobile app -- the input to the classifier model -- rather than to the app itself.  That is, the source input is the original feature vector for the app and the derived input is that vector with selected top-k benign features removed.  If the app was originally classified benign, and is indeed benign, the output for the source and derived inputs should be the same class, i.e., benign. But if they differ, then the app is flagged as (likely) malware. Apps originally classified as malware should retain that classification, since only top-k benign features are considered for removal. 

We trained a vanilla neural network on 3 million app samples collected from The AndroZoo dataset. We used VirusTotal and Androwarn to vet the apps and extract features that represent the app behavior, respectively. We evaluated our technique using a vanilla neural network, as a baseline model, and the RePack dataset~\cite{li2019rebooting} for testing. We demonstrate that \deceit achieves an accuracy of 94.56\% on the repackaged app dataset while consistently detecting benign applications with an accuracy of 97.77\%. The downside is a 1.82\% drop in the detection rate of legitimately benign apps. The benefit is its security against repackaging attacks. Thus, we interpret the 1.82\% difference as the cost of security for improved malware classification. 

Our open-source implementation, models, and complete training \& testing datasets are available at https://github.com/Programming-Systems-Lab/DECEIT. 
In future work, we will investigate applying \deceit to other learning methods such as random forest and SVM.  We also plan further study of \deceit's effectiveness when attackers have full knowledge of the defense, e.g., it may be possible to infer the specific benign features to be reduced, from knowledge of the training set or the model itself. 



\bibliographystyle{IEEEtran}
\bibliography{MET_2021}




\end{document}